\PassOptionsToPackage{table,xcdraw}{xcolor}
\documentclass[table,twocolumn]{aastex61}
\usepackage[table,xcdraw]{xcolor}
\usepackage{amstext}
\usepackage{amsmath}
\usepackage{amssymb}
\usepackage{apjfonts}
\usepackage{graphicx}
\usepackage{tikz}
\usepackage{float}
\usetikzlibrary{decorations.pathreplacing,angles,quotes}

\newcommand{\beqar}{\begin{eqnarray}}
\newcommand{\eeqar}{\end{eqnarray}}

\newcommand{\vinf}{v_\infty}
\newcommand{\rhoinf}{\rho_\infty}
\newcommand{\mach}{\mathcal M}

\newcommand{\Rs}{R_{\star}}

\newcommand{\g}{\gamma}

\newcommand{\rac}{R_{\rm a,c}}
\newcommand{\rastar}{R_{\rm a,{\star}}}

\newcommand{\rperp}{R_\perp}
\newcommand{\rc}{R_{\rm c}}

\newcommand{\mdotHLstar}{\dot{M}_{\rm HL,\star}}

\newcommand{\mdotstar}{\dot{M}_\star}
\newcommand{\mdotc}{\dot{M}_{\rm c}}
\newcommand{\mdotone}{\dot{M}_{N=1}}

\newcommand*\tave[1]{\langle{#1}\rangle_t}
\newcommand*\nave[1]{\langle{#1}\rangle_N}
\newcommand*\ntave[1]{\langle{#1}\rangle_{N,\,t}}

\newcommand{\beq}{\begin{equation}}
\newcommand{\eeq}{\end{equation}}

\definecolor{andrea}{HTML}{FF376A}

\definecolor{nick}{HTML}{1e9b14}

\begin{document}

\title{Bondi-Hoyle-Lyttleton Accretion onto Star Clusters}

\correspondingauthor{Nicholas Kaaz}
\email{nomahen@ucsc.edu}

\author{Nicholas Kaaz}
\affiliation{Department of Astronomy \& Astrophysics, University of California, Santa Cruz, CA 95064, USA}
\affiliation{Niels Bohr Institute, University of Copenhagen,Blegdamsvej 17, DK-2100 Copenhagen, Denmark}

\author[0000-0003-3062-4773]{Andrea Antoni}
\affiliation{Department of Astronomy, University of California, Berkeley, CA 94720, USA}
\affiliation{Department of Astronomy \& Astrophysics, University of California, Santa Cruz, CA 95064, USA}

\author[0000-0003-2558-3102]{Enrico Ramirez-Ruiz}
\affiliation{Department of Astronomy \& Astrophysics, University of California, Santa Cruz, CA 95064, USA}
\affiliation{Niels Bohr Institute, University of Copenhagen,Blegdamsvej 17, DK-2100 Copenhagen, Denmark}

\begin{abstract} 
An isolated star moving supersonically through a uniform gas accretes material from its gravitationally-induced wake.   The rate of accretion is set by the accretion radius of the star and is well-described by classical Bondi-Hoyle-Lyttleton theory.  Stars, however, are not born in isolation.  They form in clusters where they accrete material that is  influenced by  all the stars in the cluster. We perform three-dimensional hydrodynamic simulations of clusters of individual accretors embedded in a uniform-density wind in order to study how the accretion rates experienced by individual cluster members are altered by the properties of the ambient gas and the cluster itself. We study accretion as a function of  number of cluster members, mean separation between them, and size of their individual accretion radii. We determine the  effect of these key parameters on the aggregate and individual accretion rates, which we compare to analytic predictions. We show that when the accretion radii of the individual objects in the cluster substantially  overlap, the surrounding gas is effectively accreted into the collective potential of the cluster prior to being accreted onto the individual stars. We find that individual cluster members can  accrete drastically   more  than  they  would  in  isolation, in particular when the flow is able to cool efficiently. This effect could potentially modify the luminosity of accreting compact objects in star clusters and could lead to the rejuvenation of young star clusters as well as globular clusters with low-inclination and low-eccentricity. 
\end{abstract}

%%%%%%%%%%%%%%%%%%%%%%%%%%%%%%%%%%%%%%%%%%%%%%%%
%%%%%%%%%%%%%% 1: INTRODUCTION %%%%%%%%%%%%%%%%%
%%%%%%%%%%%%%%%%%%%%%%%%%%%%%%%%%%%%%%%%%%%%%%%%

\section{Introduction}
\label{sec:intro}
Classical Bondi-Hoyle-Lyttleton (BHL) theory describes accretion onto a single point mass embedded in a uniform-density background gas  \citep{hoyle_lyttleton_1939,bondi_hoyle_1944,bondi_1952}. This scenario has been studied extensively analytically \citep{bisnovatyi_1979,1999A&A...347..901F,horedt_2000,2005A&A...435..397F} and has been confirmed numerically \citep{hunt_1971,shima_1985,ruffert_1994a,ruffert_1994b,ruffert_1994c,blondin_2012,2015MNRAS.454.2657E,2018MNRAS.478..995B}.  

While the BHL formalism can be used to describe accretion onto solitary stars moving relative to a background gas, it remains unclear how accretion changes if the stars  are members of a cluster.  Stars for the most part do not form in isolation. Instead, they form in clusters \citep{2018arXiv181201615K}, and in these  environments stars can have unsubtle effects on one another as well as on their surrounding gas. 

Numerical studies of BHL accretion have been restricted to single accretors\footnote{The term {\it accretor} refers to a generalized code construct that has the properties of (i) accreting ambient gas that falls within its radius and (ii) exerting an individual gravitational potential.} or, in rare cases, to accreting binaries \citep{bode_2012, giacomazzo_2012, farris_2010, shi_2015}.  The study of BHL accretion onto a cluster of discrete point masses has not previously been performed.
Instead, clusters have traditionally been modeled by smooth, continuous potentials whose exact form depends on the density distribution of stars within the system. These \textit{core potentials}, as they are usually referred to, are generally constructed with  specific functional forms that are observationally-motivated \citep{plummer_1911,king_1966,hernquist_1990}.  Core potentials have the advantage of being more easily described analytically and numerically \citep{lin_2007,naiman_2009,naiman_2011}.  Since core potentials represent the limiting case of a cluster with an infinite number of accretors, they are only approximate models for stellar systems with an enormous number of stars, such as nuclear clusters or globular clusters.  Lighter  stellar systems, such as open clusters, behave as discontinuous potentials, making the core potential approximation even less appropriate.  Studies of accretion onto core potentials also do not directly measure the accretion of material onto individual stars within the cluster.  This makes simulations of discretized potentials especially advantageous, since they allow the specific accretion rates of the cluster members  themselves to be studied from first principles.

In this work, we extend the analytical and numerical framework of classical BHL accretion to star clusters. We perform three-dimensional hydrodynamic simulations of clusters of accretors embedded in a uniform-density wind in order to study how the accretion rates experienced by the individual cluster members are altered by the properties of the ambient gas and by the cluster itself. We study accretion as a function of the number of cluster members, the mean separation between them and the size of their individual accretion radii.  We also consider the accretion rates realized by individual accretors as a function of their position within the cluster.  

This paper is organized as follows.  In Section \ref{sec:background}, we briefly review the characteristic scales of BHL accretion and then extend them to the study of multiple accretors. In Section \ref{sec:procedure}, we provide specific details about the simulation setup and tabulate the full parameter space of the simulations we have performed. In Section \ref{sec:results}, we present the flow and accretion properties of our simulations.  We then present the dependence of the accretion rates  on the properties of the cluster, the total number of accretors and the equation of state of the gas.  In Section \ref{sec:discussion}, we provide a summary, consider the implications of our results to relevant astrophysical settings and present a brief discussion on the dependence of individual accretion rates on location within a cluster. 

%%%%%%%%%%%%%%%%%%%%%%%%%%%%%%%%%%%%%%%%%%%%%%%%
%%%%%%%%%%%%%% 2: BACKGROUND %%%%%%%%%%%%%%%%%%%
%%%%%%%%%%%%%%%%%%%%%%%%%%%%%%%%%%%%%%%%%%%%%%%%
\section{Setting the Stage}
\label{sec:background}
The original Hoyle-Lyttleton (HL) problem \citep{hoyle_lyttleton_1939,hoyle_lyttleton_1940a,hoyle_lyttleton_1940b} considers the accretion by a point mass moving supersonically through a uniform density medium.  The upstream flow is characterized by a Mach number $\mathcal{M}=v_{\infty}/c_{\infty}$ and a gas density ${\rho}_{\infty}$. Here $c_{\infty}$ is the sound speed and $v_{\infty}$ is the speed of the point mass relative to the gas. 

In the supersonic regime, the accretion radius 
\begin{equation}
R_{\rm a} = \frac{2GM}{\vinf^2},
\label{accretion_radius}
\end{equation}
characterizes the length scale of gravitational influence of the point mass on the gas. This length scale sets the cross section through which matter flowing past the point mass is captured into a downstream wake and is subsequently accreted \citep{edgar_2004}.

The HL accretion rate is defined as the flux of material with impact parameter less than $R_{\rm a}$,
\begin{equation}
\dot{M}_{\rm HL} = \pi R_{\rm a}^2 \vinf \rhoinf = \frac{4\pi G^2M^2\rhoinf}{\vinf^3}.
\label{mdot_hl}
\end{equation}

A key goal of this paper is to learn how the prediction of HL must be modified to describe the flow in and around a star cluster hosting $N$ accretors, each of mass $M_\star$. To that end, it is necessary to make the distinction between the accretion radius of an individual cluster member, denoted 
\begin{equation}
\rastar=\frac{2G M_\star}{\vinf^2},
\end{equation}
and the accretion radius of the entire cluster, denoted 
\begin{equation}
\rac =\frac{2G M_{\rm c}}{\vinf^2},
\end{equation}
where $M_{\rm c}=NM_\ast$ is the total mass of the cluster.

Within a stellar cluster, the characteristic accretion radius, $R_{\rm a,{\star}}$, and  the mean separation between cluster members,
\begin{equation}
\rperp =R_{\rm c} N^{-1/3},
\end{equation}
determine whether individual accretors  are able to influence one another. Here, $R_{\rm c}$ is the cluster's physical radius. In this context, traditional BHL flow will take place around each individual accretor when   $\rperp \gg \rastar$. That is, the separation between accretors is much larger than their accretion radii and each accretor can be considered to accrete independently. Then the mean accretion rate of the individual cluster members is roughly
\begin{equation}
\langle\dot{M_\star}{\rangle}_{\rm ind} \approx \dot{M}_{\rm HL,\star} = \pi \rastar^2 v_{\infty} {\rho}_\infty = \frac{4\pi G^2M_\star^2{\rho_{\infty}}}{v_{\infty}^3},
\end{equation}
where $\dot{M}_{\rm HL,\star}$ is the rate at which a cluster member is expected to accrete in isolation, giving a total accretion rate for the cluster of
\begin{equation}
\dot{M}_{\rm c,ind} \approx \sum_{i=1}^N  \langle\dot{M_\star}{\rangle}_{\rm ind}\approx  N \dot{M}_{\rm HL,\star}.
\end{equation}
When $\rperp \gg \rastar$, we  expect the flow structure to be composed of individual, detached bow shocks around each cluster member.

The key difference that this work will emphasize is the role of $R_\perp/R_{\rm a,{\star}}$ in shaping the flow around the individual accretors when $R_\perp/R_{\rm a,{\star}}\lesssim 1$. In this case, the accretion radii of the individual accretors substantially overlap, and one naively could consider the cluster to accrete in aggregate such that
\begin{eqnarray}
\dot{M}_{\rm c, col}\approx \dot{M}_{\rm HL,c} = \pi \rac^2 \vinf \rhoinf = \frac{4\pi G^2(N M_\star)^2 \rhoinf}{\vinf^3}. 
\label{eq:mc}
\end{eqnarray}
Equation~\ref{eq:mc} can be rewritten as
\begin{equation}
\dot{M}_{\rm c, col} \approx N^2 \dot{M}_{\rm HL,\star}=N \dot{M}_{\rm c, ind},
\end{equation}
so that the mean accretion rate of an individual cluster member is 

\begin{equation}
  \langle\dot{M_\star}{\rangle}=\begin{cases}
    \dot{M}_{\rm HL,\star}\;\;\;\;\; {\rm when}\;\rperp\gg\rastar\\
    N \dot{M}_{\rm HL,\star}\;\; {\rm when}\;\rperp \lesssim \rastar.
  \end{cases}
\end{equation}

This suggests that when the entire cluster is able to  gravitationally focus ambient gas, each individual cluster member could accrete significantly more than it would in isolation \citep{lin_2007}. In the limit of $R_{\rm a,c}>R_{\rm c}$, the flow structure is expected to be composed of a single, coherent bow shock \citep{naiman_2009,naiman_2011}.

This simple analysis, however,  considers the gravitational capture of the ambient gas by the entire cluster and neglects the three-dimensional distribution of the individual accretors. Though this gas is by construction bound to the cluster, the individual cluster members are not necessarily able to accrete the gas efficiently. In general, we expect the mean individual accretion rates to scale as
\begin{equation}
\langle\dot{M_\star}\rangle = N^{\alpha}\dot{M}_{\rm HL,\star}
\label{mdot_alpha}
\end{equation}
where $0 \lesssim \alpha \lesssim 1$. The accretion rate depends not only on $N$, but also (through $\alpha$) on $R_{\perp}/R_{\rm a,\star}$. In this prescription, $\alpha = 0$ gives the limiting case when $R_{\perp} \gg R_{\rm a,{\star}}$ while $\alpha = 1$ represents the  opposite limit when $R_{\perp} \ll R_{\rm a,{\star}}$. The goal of this paper is to perform numerical simulations based on the phase space of  the physically motivated flow parameters derived in this section in order to empirically determine the relevant  values for $\alpha$. 

%%%%%%%%%%%%%%%%%%%%%%%%%%%%%%%%%%%%%%%%%%%%%%%%
%%%%%%%%%%%%%%%% 3: PROCEDURE %%%%%%%%%%%%%%%%%%
%%%%%%%%%%%%%%%%%%%%%%%%%%%%%%%%%%%%%%%%%%%%%%%%
\section{Methods}
\label{sec:procedure}

Classical numerical studies simulate HL accretion by fixing an accreting point mass at the origin of the computational domain and sending an (initially) uniform-density flow past the object in a ``wind-tunnel'' \citep{blondin_2012}. To study the effect of a cluster of point masses on the upstream flow, we modify this approach by embedding a cluster of $N$ accretors in the wind tunnel. In this section, we describe our numerical methods in detail. 

All simulations are performed using version 4.5 of the grid-based adaptive mesh refinement hydrodynamics code FLASH \citep{fryxell_2000}. We employ FLASH's directionally split hydrodynamics solver using the direct Eulerian version of the piecewise-parabolic method (PPM), which is described in \cite{colella_woodward_1984}. We make use of
a gamma-law equation of state, and in general we use $\gamma=5/3$. A $\gamma=1.1$ equation of state  is also used here to simulate the properties of the flow when significant cooling occurs in the gas, which we denote as quasi-isothermal. This work extends a  tradition of numerical study of HL accretion in our group \citep{naiman_2009,naiman_2011,2015ApJ...803...41M,2015ApJ...798L..19M,2017ApJ...845..173M,2017ApJ...838...56M}. The reader is referred to \citet{2015ApJ...803...41M} for a detailed review  of previous numerical work that summarizes results with respect to flow stability as it depends on sink size, Mach number, equation of state and geometry.

\subsection{Boundary and Initial Conditions}
We initialize a three-dimensional Cartesian grid with uniform ambient gas properties $P_{\infty}$ and ${\rho}_{\infty}$ for the pressure and density, respectively. The gas is initially moving in the $+x$ direction with uniform velocity $v_{\infty}$, and is replenished at the $-x$ boundary every time step to simulate the incoming flow. The $y$ and $z$ boundaries are chosen so that the bow shock formed by the wind is fully  contained within the domain. We use HL units, setting the cluster accretion radius $R_{\rm a,c} = 1$ and setting the wind velocity $v_{\infty} = 1$. 
 
\subsection{Accretion}
The $N$ cluster members are distributed uniformly, using rejection sampling to fix their separation to values near $R_{\perp}$.  The center of mass of the cluster is placed at the coordinate origin. Each cluster member is represented by a point mass that is surrounded by an absorbing boundary characterized by a sink radius $R_\star$ and whose position is fixed to the grid. The accretion algorithm samples the gas properties of cells within $R_{\star}$ and floors the values of density and pressure within each cell. The removed mass is recorded, but not added to the sink under the assumption that the mass of the accumulated gas is much less than $M_{\star}$. The sink boundary is softened, such that more mass is accreted in the center of the sink than  near the surface of the sink boundary. This prevents numerical oscillations and helps keep the time-step from becoming too small. The density and pressure of cells within the sink boundary are not allowed to fall below $15\%$ of their ambient values, ${\rho}_{\infty}$ and $P_{\infty}$, respectively.

\subsection{Gravity}
 The gravitational potential  of the cluster is fully discretized, with each accretor having equal mass. The self-gravity of the gas is neglected.  The gravitational potential of the individual accretors is turned on gradually over one HL crossing time of the cluster, $R_{\rm a,c}/v_{\infty}$.  Additionally, the gravitational acceleration of each point mass is modified by a softening parameter similar to that of \cite{ruffert_1994a}, where $\epsilon$ is the softening parameter, $\delta$ is the smallest cell size of the simulation, and $r_i$ is the distance of the cell from the potential of the $i$-th accretor
\begin{equation}
\boldsymbol{\vec{\textbf{a}}}_{\rm g} =\sum_{i=1}^N \frac{-GM_{\star}}{r_i^2 + {\epsilon}^2{\delta}^2{\rm exp}(-r_i^2/{\epsilon}^2{\delta}^2)}{\hat{\textbf{r}}}_{\rm i}.
\end{equation}
This  prevents the flow from diverging near the boundary of the individual accretors and ensures that the time step does not become too small. 

\subsection{Resolution}
The length scales of the physics involved in our simulations span several orders of magnitude. At the largest scale, it is necessary to capture the full width of the bow shock, as otherwise we would strip pressure support away from the mass accumulation at the core of the cluster. It is also critical to resolve the core of the cluster, at scales $\approx R_{\rm a,c}$; the vicinity of each accretor, at scales $\approx R_{\rm a,{\star}}$; and,  at the smallest scales, the boundary of the sink radius itself, at scales $\approx R_{\star}$. For the purpose of computational efficiency, we employ an adaptive mesh refinement routine. We refine preferentially on regions of higher internal energy and on regions that are near the individual accretors.

It is well-known that both the resolution and the radius of the sink affect the accretion rates realized in HL simulations \citep{ruffert_1994a,blondin_2012}. To control for this, we hold the ratio  ${\delta}:R_{\star}:R_{\rm a,{\star}}$ constant across simulations. In all simulations, we fix the cluster mass $M_{\rm c} = 1/2G$ (such that $R_{\rm a,c} = 1$), which means that as we increase $N$, the total mass is effectively  split  into lighter components. To facilitate comparison, we normalize all results to the accretion rate of a single accretor in isolation, $\dot{M}_{N=1}$, at the same resolution and sink size; that is, the same values of $\delta:R_{\star}:R_{\rm a,{\star}}$. In general, $\dot{M}_{N=1} \approx \dot{M}_{\rm HL,\star}$, although the exact value of $\dot{M}_{N=1}$ depends on resolution and sink size. As a consistency check, we performed tests with $R_{\star}/\delta$ varying from 12.5 to 4  with no  appreciable difference in the normalized accretion rates.

\begin{table}
\centering
{\caption{All simulations presented in this work are listed. All simulated clusters are distributed uniformly. The computational volume is the same in all simulations: $[x_{\rm min},x_{\rm max}]=[-12.0,20.0]$ $R_{\rm a,c}$, $[y_{\rm min},y_{\rm max}]=[-22.4,22.4]$ $R_{\rm a,c}$, $[z_{\rm min},z_{\rm max}]=[-22.4,22.4]$ $R_{\rm a,c}$.}
\label{sim_table}}
{\begin{tabular}{lcccc}
\hline
$N$ & $\rperp\ [\rastar]$ & $\gamma$ & $R_{\rm c}\ [\rac]$ & $\mathcal{M}$ \\ \hline\hline
4   & 0.2                 & 5/3      & 0.08 & 2.0 \\ 
4   & 0.5                 & 5/3      & 0.20 & 2.0 \\ 
4   & 2                   & 5/3      & 0.79 & 2.0 \\ \hline
16  & 0.2                 & 5/3      & 0.03 & 2.0 \\ 
16  & 0.5                 & 5/3      & 0.08 & 2.0 \\ 
16  & 2                   & 5/3      & 0.31 & 2.0 \\ \hline
16  & 0.5                 & 5/3      & 0.08 & 1.5 \\
16  & 0.5                 & 5/3      & 0.08 & 3.0 \\ \hline
32  & 0.2                 & 5/3      & 0.02 & 2.0 \\ 
32  & 0.5                 & 5/3      & 0.05 & 2.0 \\ 
32  & 2                   & 5/3      & 0.20 & 2.0 \\ \hline
4   & 0.5                 & 1.1      & 0.20 & 2.0 \\ 
16  & 0.5                 & 1.1      & 0.08 & 2.0 \\ 
32  & 0.5                 & 1.1      & 0.05 & 2.0 \\ \hline\end{tabular}}
\end{table}
\begin{figure*}
\includegraphics[width=\textwidth]{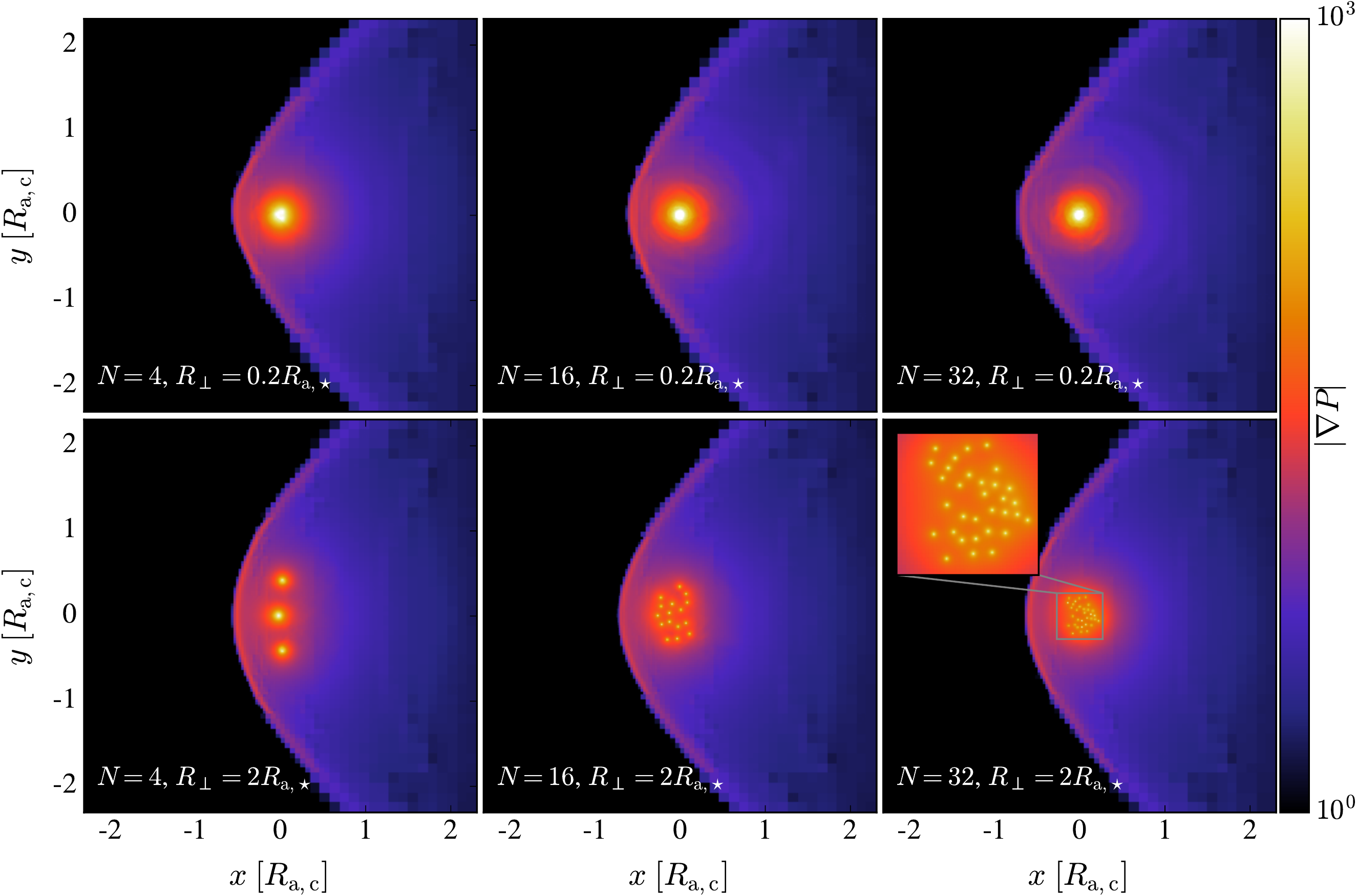}
\caption{The projection of the large-scale flow structure is depicted for the $\rperp = 0.2$ $\rastar$ (top row panels) and $\rperp = 2$ $\rastar$ (bottom row panels) simulations for $N = 4$, $16$, and $32$ with $\gamma = 5/3$. The magnitude of the pressure gradient, in units of $\rhoinf \vinf^2/\rac$, is mapped at time $32$ $\rac/\vinf$. In all simulations a single, coherent bow shock is formed after a time $\approx 10$ $R_{\rm a,c}/v_{\infty}$ and a steady density enhancement envelops the cluster, consistent with studies of core potentials \citep{naiman_2009,naiman_2011}.\\}
\label{flow_structure_adiabatic}
\end{figure*}

\subsection{Simulation Parameters}\label{subsec:sp}
We perform fourteen simulations (Table \ref{sim_table}). In all calculations we fix the softening parameter ($\epsilon=7$, which serves only to prevent spurious velocity divergences near accretor boundary and its exact value has a negligible effect on the flow), the cluster mass ($M_{\rm c}$ =$1/2G$, which is done for numerical convenience), the parameters controlling the resolution ($R_{\star}/\delta=4$, $R_{\star}/R_{{\rm a,}\star}=0.05$) and the dimensions of the computational domain (see Table \ref{sim_table} caption). In most simulations we fix the Mach number of the incident wind to  be $\mathcal{M}=2$. The dependence on Mach number on the flow is discussed in Section~\ref{sec:mach}.  ~For all simulations, a steady state solution is achieved by time $\approx 10$ $R_{\rm a,c}/v_{\infty}$, and each simulation is run until a time of $32$ $R_{\rm a,c}/v_{\infty}$. An adiabatic ($\gamma = 5/3$) equation of state is used for 9 simulations and a quasi isothermal ($\gamma = 1.1$) equation of state is used for the other 3. The two key parameters that are systematically varied are the number of accretors ($N = 4,16,32$) and the mean separation between accretors ($R_{\perp}/R_{\rm a,{\star}} = 0.2,0.5,2$). Because $R_{\perp}/R_{\rm a,{\star}}$ is altered systematically  and the cluster members are assumed to be distributed uniformly, the corresponding  changes in cluster radius, $R_{\rm c}$, can be calculated using the following relation
\begin{equation}
\frac{R_{\rm c}}{R_{\rm a,c}} = \left(\frac{R_{\perp}}{R_{\rm a,{\star}}}\right)N^{-2/3}.
\label{ec:rc_scaling}
\end{equation}
The value of $R_{\rm c}$ for each simulation is reported in Table \ref{sim_table}. Equation \ref{ec:rc_scaling} implies  $\sigma_{\rm c}\propto N^{1/3}$, where $\sigma_{\rm c}$ is the velocity dispersion of our simulated clusters. For comparison, $\sigma_{\rm c}\propto N^{1/2}$ for globular clusters \citep{2005ApJ...627..203H} and $\sigma_{\rm c}\propto N^{1/4}$ for young stellar clusters \citep{2016A&A...586A..68P}.

\section{Results}
\label{sec:results}
In this section, we present the results of all simulations listed in Table \ref{sim_table}. We first discuss the macroscopic flow structure for the adiabatic simulations. Then, we consider the dependence of individual accretion rates on $\rperp / \rastar$ and on $N$. Finally, we investigate effects of different equations of state ($\gamma = 5/3$ and $1.1$) on the morphology of the flow and the accretion rates of the individual cluster members.

\subsection{Macroscopic Properties of the Flow}
\label{flow_properties}

Figure \ref{flow_structure_adiabatic} shows projection plots for our adiabatic simulations with the smallest and largest values of the mean separation, $\rperp/\rastar = 0.2$ and $2$, respectively, at a time of $t = 32$ $\rac / \vinf$. These snapshots exemplify features common to all of our adiabatic simulations.  The large-scale structure retains the key features of canonical BHL flow: supersonic flow leads to the formation of a single, coherent bow shock upstream from the cluster as material is funneled into a high-density wake downstream.  The shock is smooth and axisymmetric about the line of motion of the cluster's center of mass.

A striking feature of all of our adiabatic simulations is the existence of a finite standoff distance between the cluster and the bow shock. Because the entire cluster is always fully embedded behind the shock, the mass accumulation onto the cluster is insulated from the ram pressure of the wind. This feature, consistent with studies of core potentials \citep{naiman_2009,naiman_2011}, allows a stable density enhancement to form around the cluster members. 

As $R_{\perp}$ decreases, variations in density are smoothed out  and the cluster's core is contained within a spherically symmetric high-density region of gas.  As is apparent in the top row of Figure  \ref{flow_structure_adiabatic}, the density enhancement is highly symmetric about the cluster center of mass and individual cluster members become indiscernible.  This is in agreement with our analysis of $\alpha$ in Section \ref{sec:background}; when $\rperp \ll \rastar$, the cluster behaves increasingly as a single accretor with mass $NM_{\star}$ (Equation \ref{eq:mc}). The symmetry of the flow is, however, broken at sub-cluster scales by the presence of the individual cluster members (see inset panel in Figure~\ref{flow_structure_adiabatic}), which leads to dramatic changes to the morphology of the flow around the individual accretors.

\begin{figure}
\includegraphics[width=1.0\linewidth]{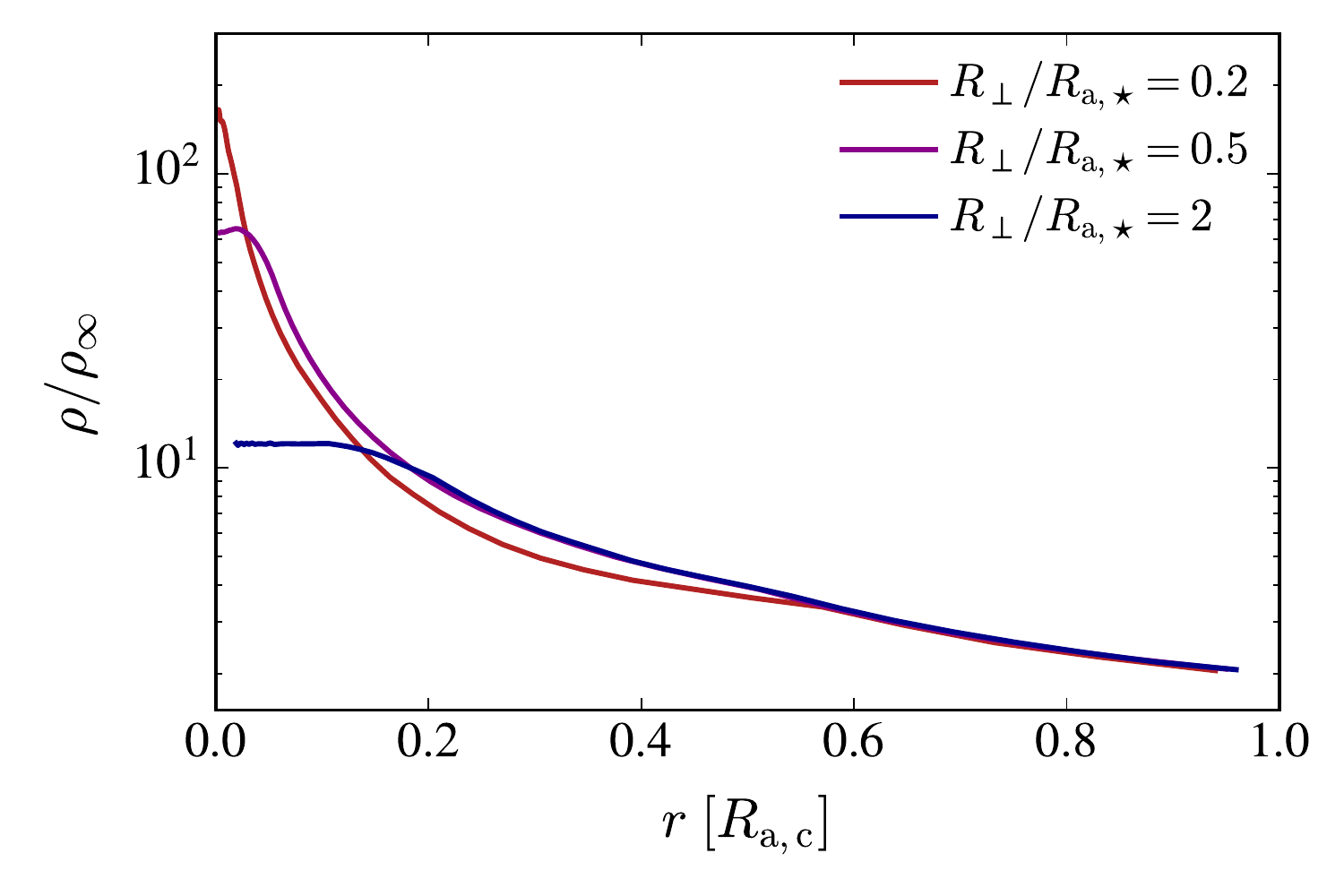}
\caption{The radial density profile is depicted for $R_{\perp} = 0.2$, $0.5$ and  $2$ $R_{\rm a,{\star}}$ simulations for $N =32$ with $\gamma = 5/3$. The radii of these clusters are $0.02$, $0.05$, and $0.20$ $\rac$, respectively.  For decreasing $R_{\rm c}/R_{\rm a,c}$ (Equation \ref{ec:rc_scaling}), the steepness of the density profile increases, which allows individual cluster members to gravitationally influence higher density material.}
\label{rho_r}
\end{figure}

In Figure \ref{rho_r}, we present the spatially-averaged radial density profiles of the $N=32$ simulations (the $\rperp/\rastar$ = $0.2$ and $2.0$ curves correspond to the right column of Figure~\ref{flow_structure_adiabatic}). In general, the density profile of the accumulated gas is expected to steepen as the cluster density increases, which in our setup is controlled by $R_{\perp}/R_{\rm a,{\star}}$ (Equation~\ref{ec:rc_scaling}). As a result, the density profile is shallower for $R_{\perp}/R_{\rm a,{\star}} = 2$  and much steeper for $R_{\perp}/R_{\rm a,{\star}} = 0.2$. This is also reflected in the pressure gradients of the envelopes in the corresponding panels of Figure \ref{flow_structure_adiabatic}. As $R_{\perp}/R_{\rm a,{\star}}$ decreases, the density of the material that is focused toward the individual cluster members increases. This allows them to accrete at systematically higher rates when compared to $\mdotone$.

%% INDIVIDUAL ACCRETION RATES
\subsection{Accretion Rates}
\label{individual_accretion}
One of the main goals of this work is to understand how the accretion rates of  individual cluster members differ from accretors in isolation. In this subsection, we investigate how these rates depend on the number of accretors, $N$, and on the mean separation between accretors, $\rperp$.

Before we proceed, let us first describe the notation we use when referring  to accretion of the individual cluster members. Our simulations measure the instantaneous rate of accretion for each cluster member at every time step. The measured value for a particular accretor is $\dot{M}_{\star, i}$. The total instantaneous accretion rate for the cluster is found by summing over all sinks and is denoted 
\begin{equation}
\mdotc=\sum_{i=1}^N \dot{M}_{\star, i}.
\end{equation}
We use angled brackets to indicate averages and include a subscript to indicate the quantity we average over, either $N$ or $t$.  For example, $\nave\mdotstar$ is the instantaneous average accretion rate over all $N$ accretors 
\beq
\nave\mdotstar = \frac{1}{N}\;\mdotc.
\eeq
A time average of that quantity is denoted by $\ntave\mdotstar$. All time averages presented in this paper are calculated over the steady-state phase, which is established at $t = 10$ $\rac/\vinf$ and lasts until the end of the simulation at $t = 32$ $\rac/\vinf$.

To facilitate comparison between all simulations, we normalize all accretion rates to $\mdotone$, which is the time averaged accretion rate computed for an isolated accretor using all of the same simulation parameters as the simulated cluster in question (namely, $\g$, $\mach$, $\Rs$, $\delta$, and the size of the computational domain). This also permits our results to be directly compared to realistic clusters (see Section~\ref{subsec:astro}).

\begin{figure}
\includegraphics[width=1.0\linewidth]{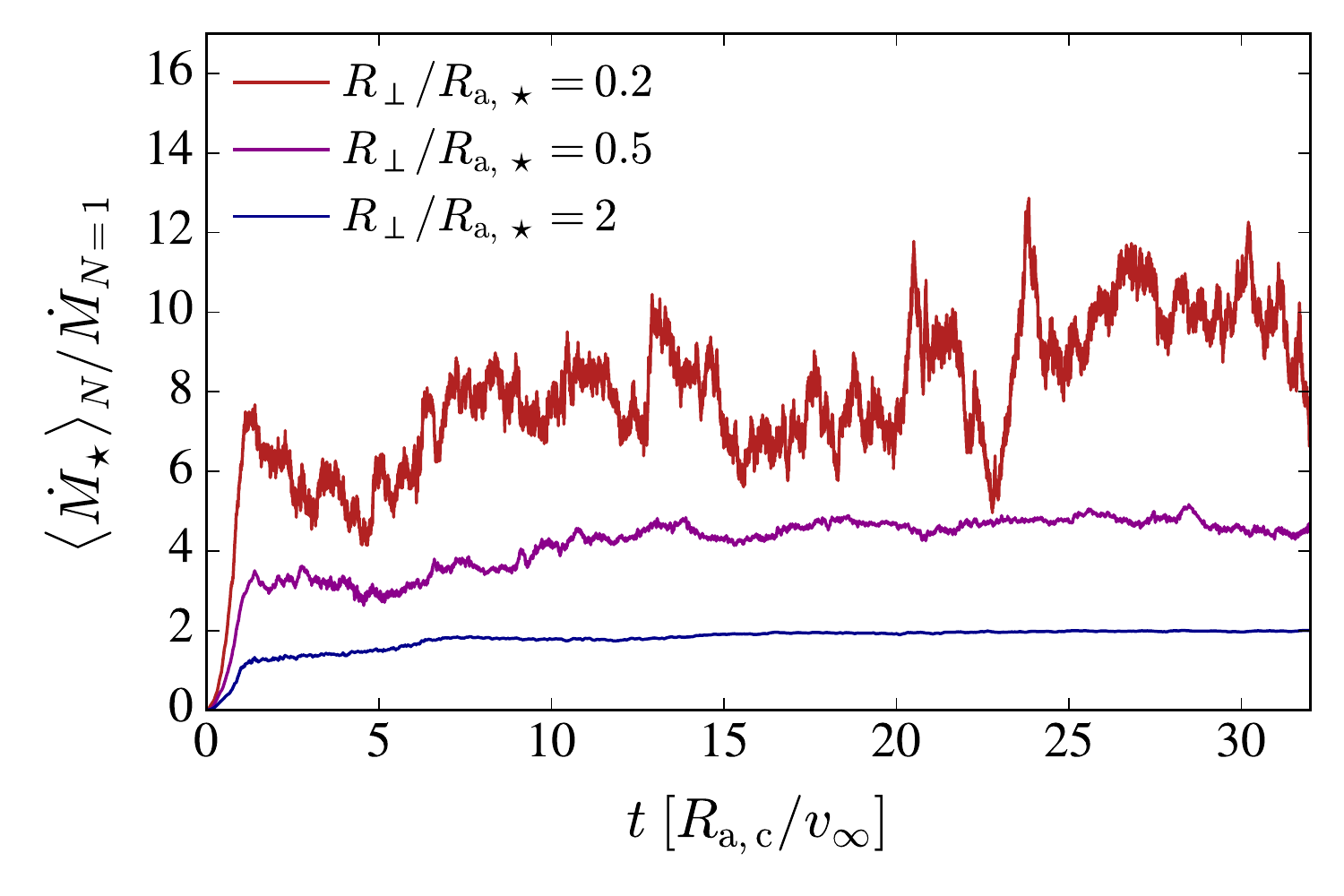}
\caption{The number-averaged accretion rates for $N = 32$, $\gamma = 5/3$ simulations are plotted as a function of time for $R_{\perp}/R_{\rm a,{\star}} = 0.2$, $0.5$ and $2$. The accretion rate is normalized to $\dot{M}_{\rm N=1} = 0.49$ $\mdotHLstar$. For decreasing $R_{\perp}$, the relative accretion rate increases and exhibits a higher degree of variability.}
\label{mdot_t_32n}
\end{figure}
\begin{figure*}
\includegraphics[width=1.0\linewidth]{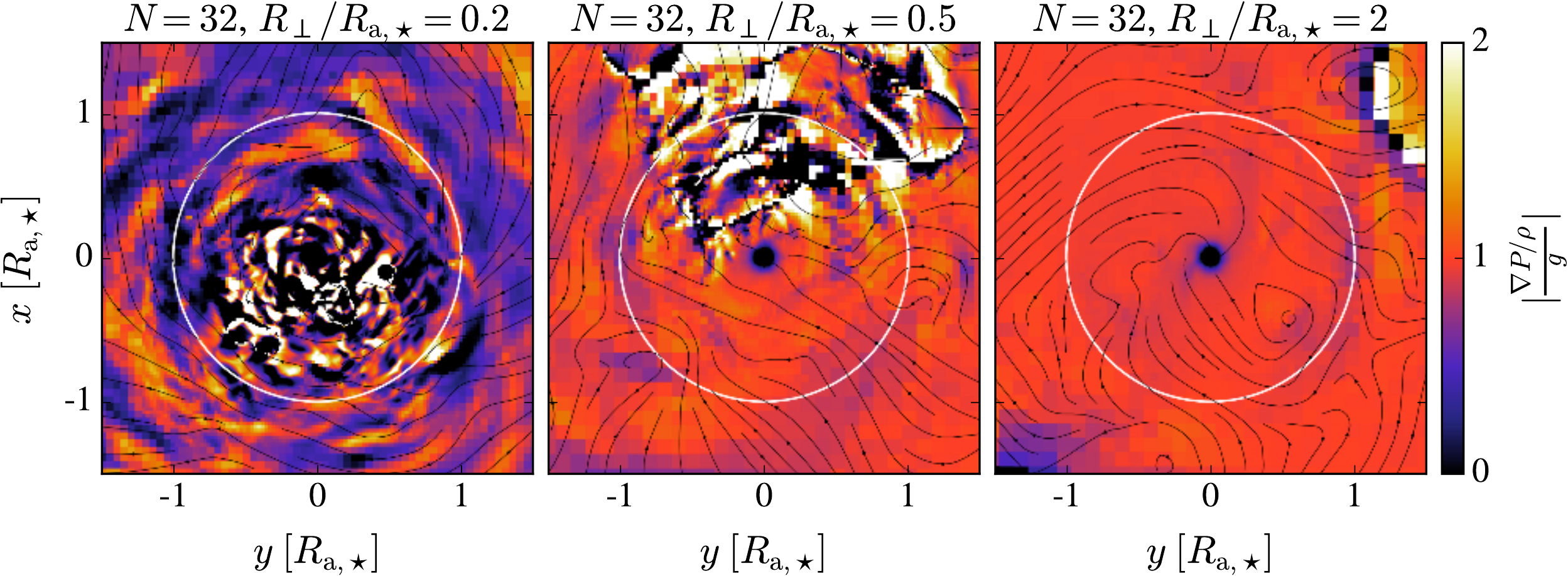}
\caption{Snapshots of the flow are shown  for the three different $N = 32$ adiabatic simulations, characterized by $\rperp/\rastar =0.2,0.5$ and 2.  The absolute value of the radial components of the pressure gradient over the gravitational force per unit density due to all objects is plotted. The white circle represents the accretion radius $\rastar$ of the individual accretors and the black arrows depict the streamlines of the flow. The pressure gradient is highly sub-dominant when $\rperp/\rastar = 0.2$, which leads to enhanced  gravitational focusing of the  flow.  As a result of the overlapping  influence of multiple members, a relatively large amount of gas is able to enter the accretion radius region of individual members.\\}
\label{mdot_pg_32n}
\end{figure*}

\subsubsection{$R_{\perp}/\rastar\;$ Dependence}
\label{sec:rperp_dependence}
Here we study average accretion rates as a function of $\rperp / \rastar$ with fixed $N$ for our adiabatic simulations.  In Figure \ref{mdot_t_32n}, we present the number-averaged accretion rates as a function of time for our adiabatic $N = 32$ simulations. As expected from the discussion of  Section \ref{sec:background}, $\nave\mdotstar$ increases with decreasing $\rperp/\rastar$. Members that are closer together (for a fixed $\rastar$) benefit from the gravitational influence of their neighbors and experience an enhanced accretion rate, on average.  

Figure \ref{mdot_t_32n} also shows that as $\rperp$ decreases, the time variability of $\nave\mdotstar$ increases.  The characteristic stirring length for turbulent motions in the high density gas that envelopes the cluster is $R_{\rm a,{\star}}$.  For the more widely separated clusters with $\rperp = 2$ and $0.5$ $R_{\rm a,{\star}}$, the cluster radius is larger than the stirring length. The different accretors within the cluster sample over all of the variations in the gas which offset one another when the average is taken. When $\rperp = 0.2$ $\rastar$, on the other hand, the cluster radius $R_{\rm c}$ becomes smaller than $R_{\rm a,{\star}}$ so the individual accretors stir the gas in aggregate. The variances in the individual accretion rates add in phase, resulting in a high variance in the number-averaged accretion rate.
 
In Figure \ref{mdot_pg_32n}, we explore the local dynamics of the flow in the immediate vicinity of an individual accretor by plotting the absolute value of the ratio of accelerations on the gas for the adiabatic $N=32$ simulations.  Here, the radial axis is defined with respect to the depicted accretor. White and yellow colors indicate regions where gas pressure dominates, red and orange colors indicate where the gas is near hydrostatic equilibrium, and black and purple regions are those in which the aggregate gravitational force dominates. 

When $\rperp/\rastar = 0.2$, the gravitational force 
dominates over the pressure gradient in most directions. This
allows a sizable number of flow lines to pierce into the
accretion radius region without being deflected by
the collisional properties of the gas. As argued above, the overlapping influence of multiple members allows the gas to be accreted into individual cluster members at drastically  enhanced rates (Figure~\ref{mdot_t_32n}). As $\rperp$ increases, the effects of gravity become weaker. For $\rperp / \rastar$ = 2 the pressure gradient establishes superiority and the flow is largely deflected away from
the accretor. As a result, the individual accretion rates are inhibited and more closely resemble those expected in isolation (Figure~\ref{mdot_t_32n}).

\subsubsection{$N$ Dependence} 
In this subsection, we consider accretion as a function of $N$ with $\rperp/\rastar$ unchanged. By fixing $\rperp/\rastar$, the overlapping influence of adjacent cluster members is set to be the same and the main changes on the individual accretion rates arise from variations in the overall density structure of the flow, which in turn  is altered by changes in the aggregate cluster potential. 

\begin{figure}
\includegraphics[width=1.0\linewidth]{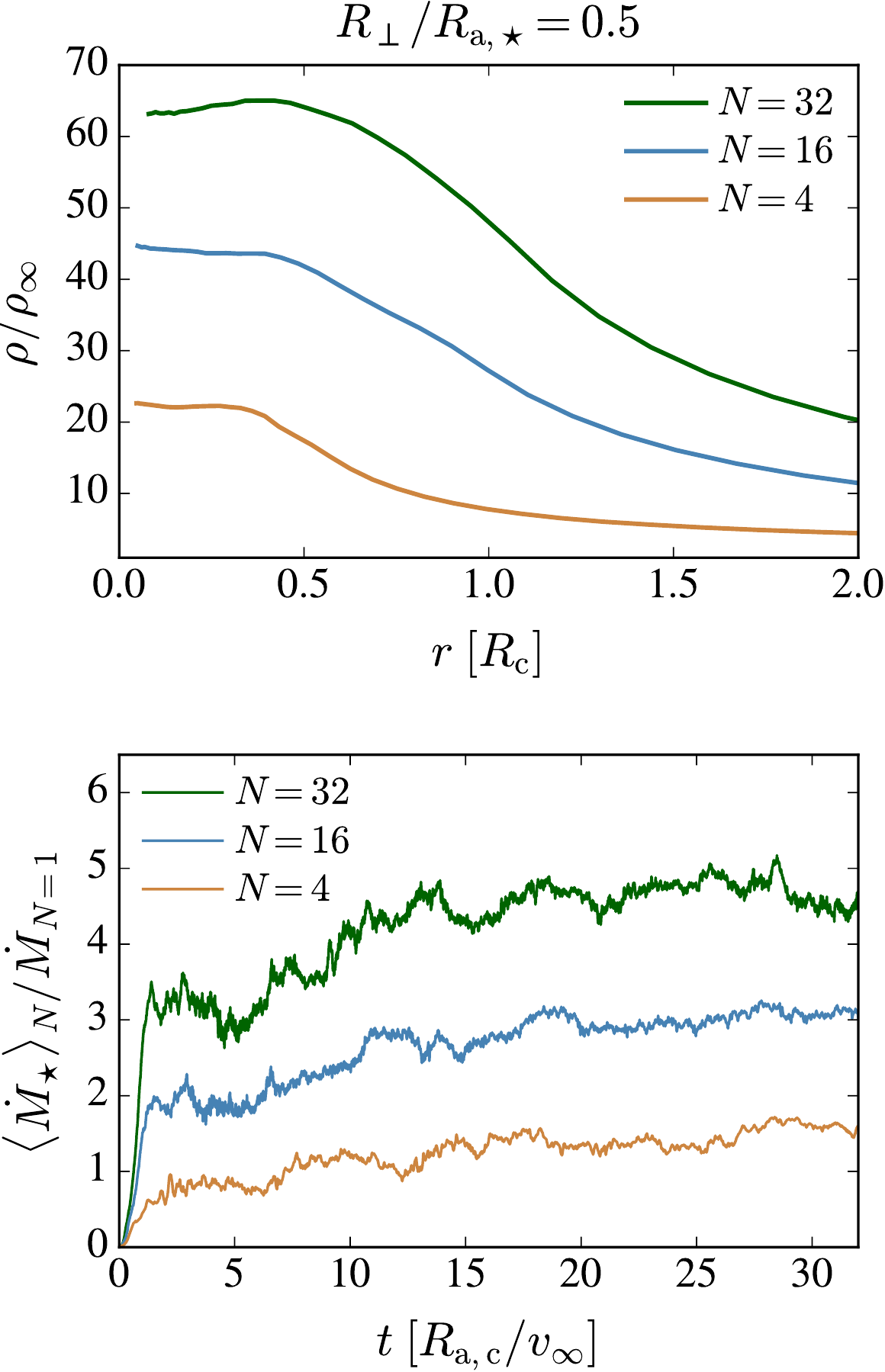}
\caption{The  radially-averaged density profiles (top panel) and  number-averaged accretion rates (bottom panel) are plotted for the $R_{\perp} = 0.5$ $R_{\rm a,{\star}}$ adiabatic simulations as a function of time. Three different values of $N$ are shown. The number-averaged accretion rate is normalized to $\dot{M}_{\rm N=1} = 0.49$ $\dot{M}_{\rm HL,{\star}}$, while the density is normalized to $\rhoinf$. For increasing $N$, the relative accretion rate increases with a similar degree of variability.}
\label{mdot_t_05rs_adiabatic}
\end{figure}

A key property of all BHL simulations is that there is a constant flux of fresh material flowing toward the central object. Interaction with this steady flow defines the density structures
seen in Figure \ref{mdot_t_05rs_adiabatic} for the adiabatic simulations with $\rperp / \rastar$ = 0.5 and $N=4, 16$ and 32. As the compactness of the cluster increases with $N$, there is a corresponding increase in the gas density near the cluster's core in order  for  ram pressure balance to be maintained with the incoming flow. The relatively higher gas densities found near the core imply that a larger amount of material is available to be accreted by the individual cluster members.  In Figure \ref{mdot_t_05rs_adiabatic}, we  also plot the number-averaged accretion rates of the individual  cluster members as a function of time. As argued above, the aggregate gravitational influence within $\rc$ yields higher density enhancements which give rise  to an enhancement of the average accretion rates of the individual cluster members  when compared to those they might experience when accreting in isolation. 

The variability in the accretion rate in Figure \ref{mdot_t_05rs_adiabatic}  is remarkably similar for different values of $N$, consistent with the stirring length arguments of Section \ref{sec:rperp_dependence} for fixed $\rperp / \rastar$. This is observed for each set of simulations with constant  $\rperp / \rastar$ and varying $N$.

\begin{figure*}
\includegraphics[width=1.0\linewidth]{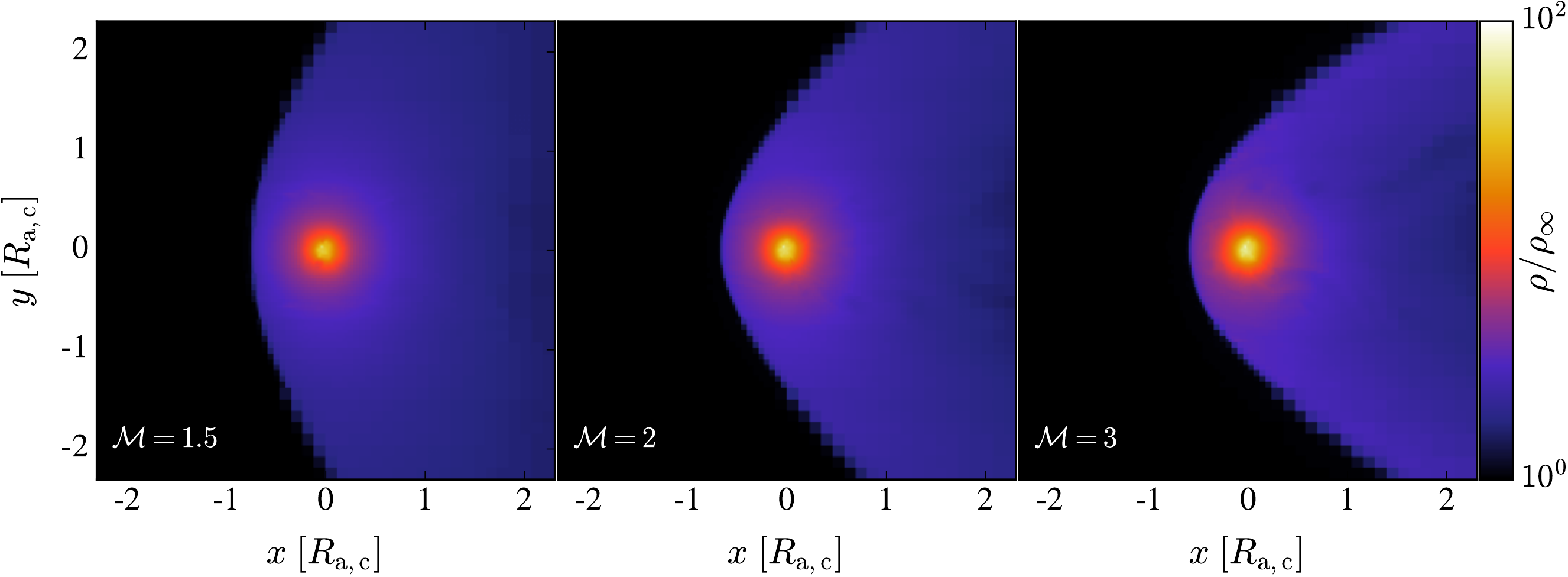}
\caption{The large-scale structure of the flow is shown for $N = 16$, $\rperp = 0.5$ $\rastar$ adiabatic simulations with varying $\mathcal{M}$. Each simulation exhibits a bow shock with standoff distance exceeding the radius of the cluster, allowing the density enhancement to remain stable. With increasing Mach number, the the standoff distance approaches the cluster radius, and the opening angle of the bow shock decreases.}
\label{flow_mach}
\end{figure*}

\begin{figure}
\includegraphics[width=1.0\linewidth]{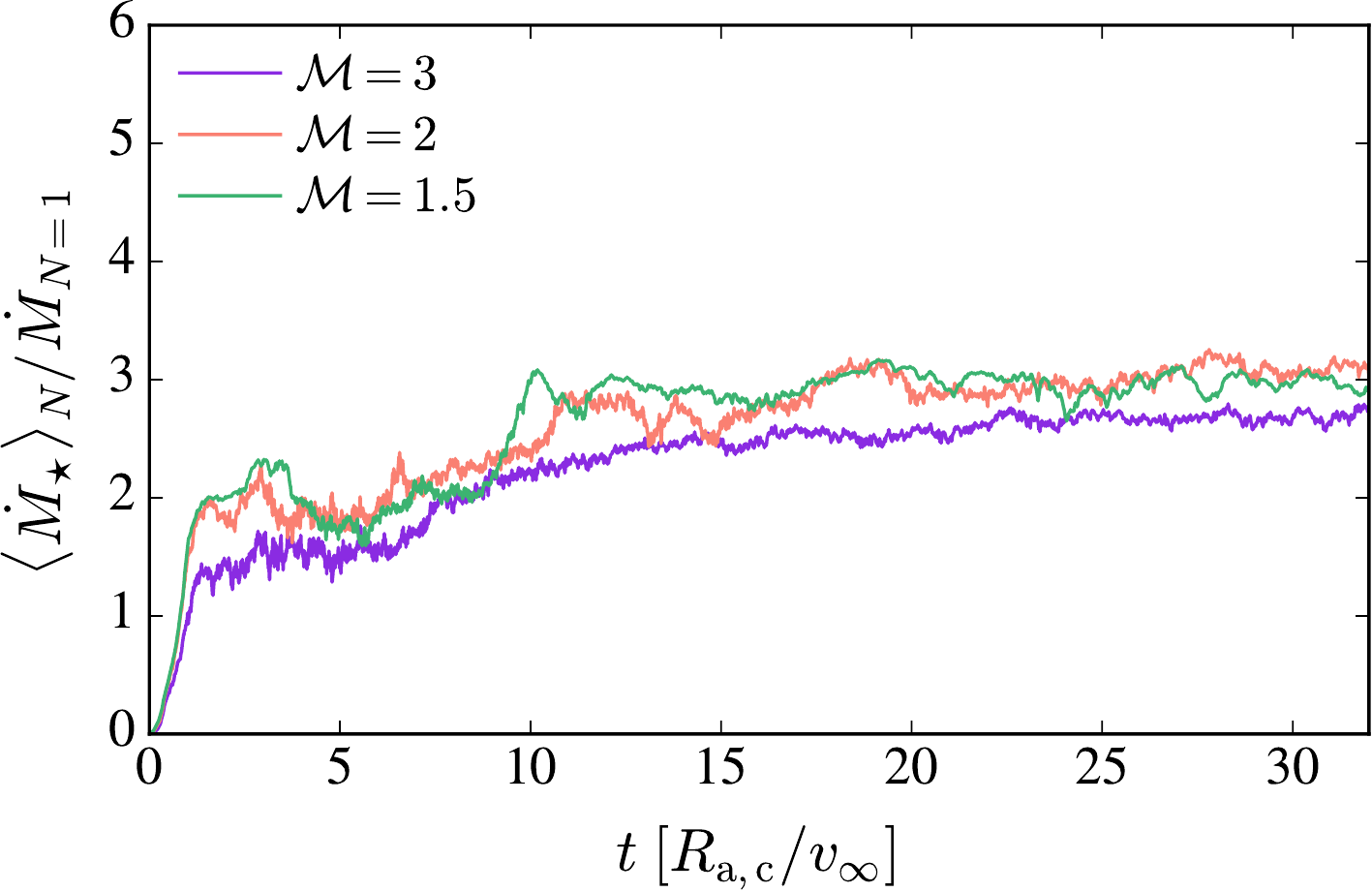}
\caption{The number-averaged accretion rates are plotted for $N=16$, $\rperp=0.5$ $\rastar$ adiabatic simulations. Three different values of $\mathcal{M}$ are shown. The number-averaged accretion rate is normalized to $\mdotone=$ ($0.42$, $0.49$, $0.68$) for $\mathcal{M}=$ ($1.5$, $2$, $3$), respectively. The relative accretion rate is enhanced similarly each choice of Mach number.}
\label{mdot_mach}
\end{figure}

\subsubsection{$\mathcal{M}$ Dependence}
\label{sec:mach}
While for the majority of this work we focus primarily on $\mathcal{M}=2$ flow, here we briefly consider the dependence of our results on $\mathcal{M}$ (namely, $\mathcal{M} = 1.5$ and $3$). In Figure \ref{flow_mach}, we examine the large-scale flow morphology for different Mach numbers. For each Mach number, a stable bow shock forms, with smaller opening angles at higher Mach numbers. The bow shock is stable because the standoff distance to the bow shock insulates the density enhancement from the ram pressure of the wind. However, with increasing $\mathcal{M}$ the separation between the bow shock and the cluster radius decreases, and at some critical Mach number the ram pressure will be able to penetrate into the cluster's core \citep{naiman_2011}. If this is the case, a stable flow structure will no longer form and the flow morphology might loosely resemble that seen in our isothermal simulations (Section \ref{subsec:results_eos}).

We explore the dependence of accretion on Mach number in Figure \ref{mdot_mach}. Plotted are our results for the normalized number-averaged accretion rates  as a function of time for different Mach numbers. Here, we normalized the number-averaged accretion rate to $\mdotone=$ ($0.42$, $0.49$, $0.68$\footnote{The $N = 1$, $\mathcal{M} = 3$ accretion rates were calculated at a higher resolution ($R_{\star}/\delta = 8$) than in our other simulations. This is because at higher Mach numbers and lower resolutions, the flow becomes highly unstable, artificially decreasing the accretion rate \citep{ruffert_1994b}. This instability is not seen at $\mathcal{M}=3$ for higher resolutions \citep{blondin_2012}, motivating our use of $R_{\star}/\delta = 8$ for this particular simulation.} for $\mathcal{M}=$ ($1.5$, $2$, $3$), respectively. We find that for different Mach numbers, the number-averaged accretion rates are remarkably similar. This is consistent with Equation \ref{mdot_alpha}, which suggests that for a given $N$ and $\rperp/\rastar$, the number-averaged accretion rate should be insensitive to the choice of $\mathcal{M}$. This relation should break down at sufficiently high Mach numbers once the ram pressure of the wind prevents the flow from forming a stable structure within the core of the cluster.

\subsection{Resolving the Connection between $\mdotstar$ and $N$}
A central motivation of this study is to determine the value of $\alpha$ as a function of $\rperp /\rastar$ in the scaling
\beq
\ntave\mdotstar = N^\alpha \mdotone
\label{eq:mdot_alpha_numerical}
\eeq
which is Equation \ref{mdot_alpha} with $\mdotHLstar$ exchanged for the numerically determined $\mdotone$.  As a reminder, the subscripts $N$ and $t$ indicate that $\ntave\mdotstar$ is an average accretion rate over all cluster members and over steady-state (to remove the time dependence).

\begin{figure}
\includegraphics[width=\linewidth]{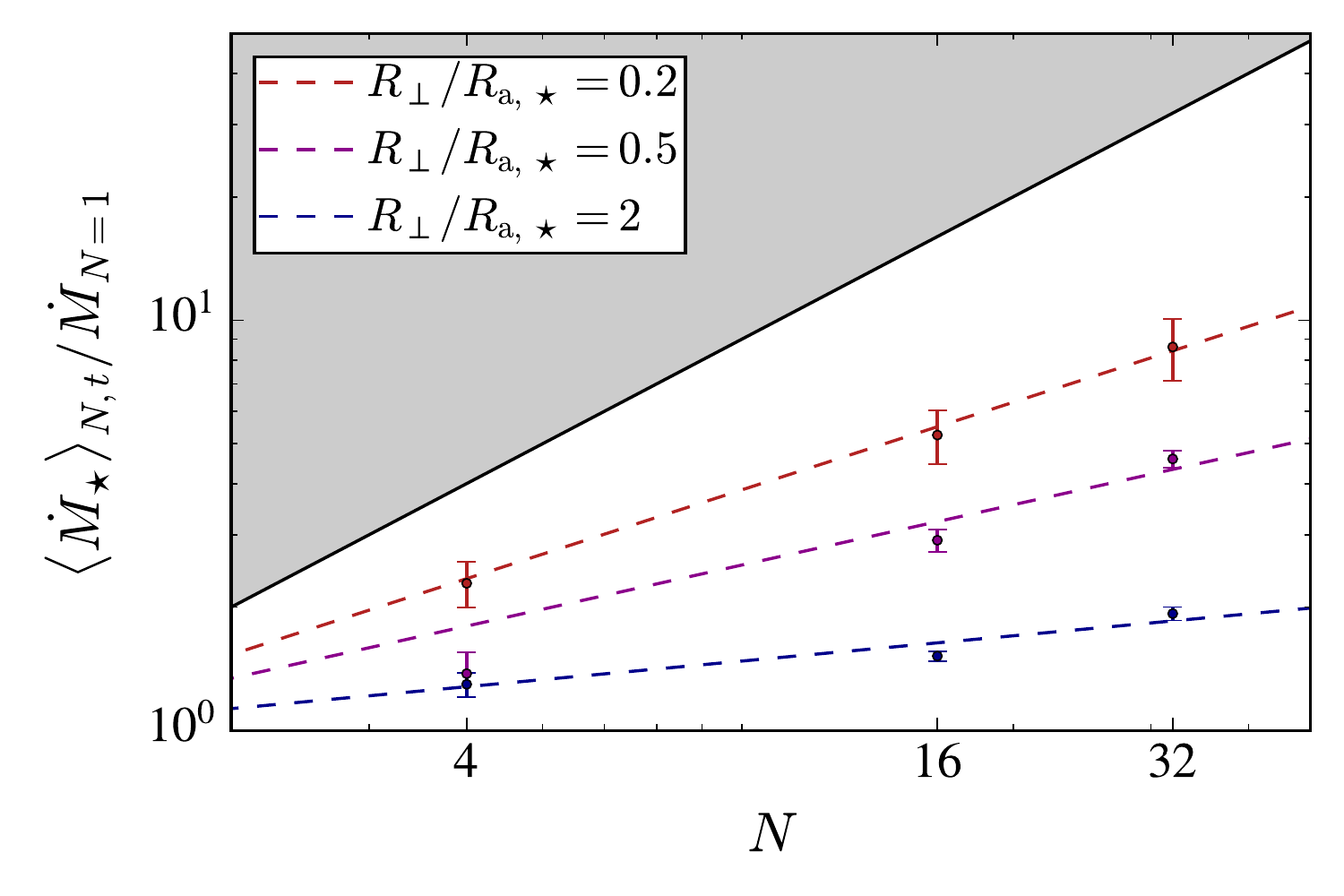}
\caption{The time- and number-averaged accretion rates are plotted for all adiabatic simulations as a function of $N$. The accretion rate is constrained by the analytic prediction in Equation \ref{mdot_dense} with $\alpha = 1$ (solid black line). Equation \ref{eq:mdot_alpha_numerical} is fit to the data for $R_{\perp}/R_{\rm a,{\star}} = 0.2$, $0.5$ and $2$ and  $\alpha$ is found to be $0.62\pm0.04$, $0.42\pm0.01$ and $0.18\pm0.01$, respectively. As the overlap between the accretion radii of the individual  cluster members increases, the relation of $\ntave\mdotstar$ with $N$  steepens although  remains sub-linear.}
\label{mdot_scale_ad}
\end{figure}

In Figure \ref{mdot_scale_ad}, we plot $\ntave\mdotstar$ in units of $\mdotone$ as function of $N$ for all adiabatic simulations.  Each dashed line represents our fit to the data for a fixed value of $\rperp/\rastar$. Fitting these data points to Equation~\ref{eq:mdot_alpha_numerical} yields  $\alpha = 0.62\pm0.04$, $0.42\pm0.01$ and $0.18\pm0.01$ for  $\rperp/\rastar = 0.2$, $0.5$ and $2$, respectively. That is, the dependence on $N$ increases as the accretion radii of the individual  cluster members more substantially overlap.  Indeed,  we find that $0 < \alpha < 1$ for all values of $\rperp/\rastar$, as expected from Section \ref{sec:background}. These results are also consistent with the limit $\alpha \rightarrow 1$ as $R_{\perp} \rightarrow 0$, with the relation always remaining sub-linear. This is because  the individual members are not able to accrete the gas efficiently\footnote{We also note that the relation is less well-defined when $N$ is small and the overall cluster potential is less smooth.}, which is due to the collisional nature of the flow.  In the $\gamma=5/3$ case, the adiabatic build-up of pressure is significant and, without a
means to dissipate its internal energy, the flow remains in
large part pressure supported. As a result, the flow is largely deflected away from the individual accretors, and a substantial fraction of the accumulated gas is advected away. When cooling is effective, as is the case for a young cluster in a dense galaxy merger environment \citep{2005AJ....130.2104W,naiman_2011}, we expect the individual accretion rates to increase  due to the loss of pressure support. It is to this issue  that we now turn our attention. 

\begin{figure*}
\includegraphics[width=\linewidth]{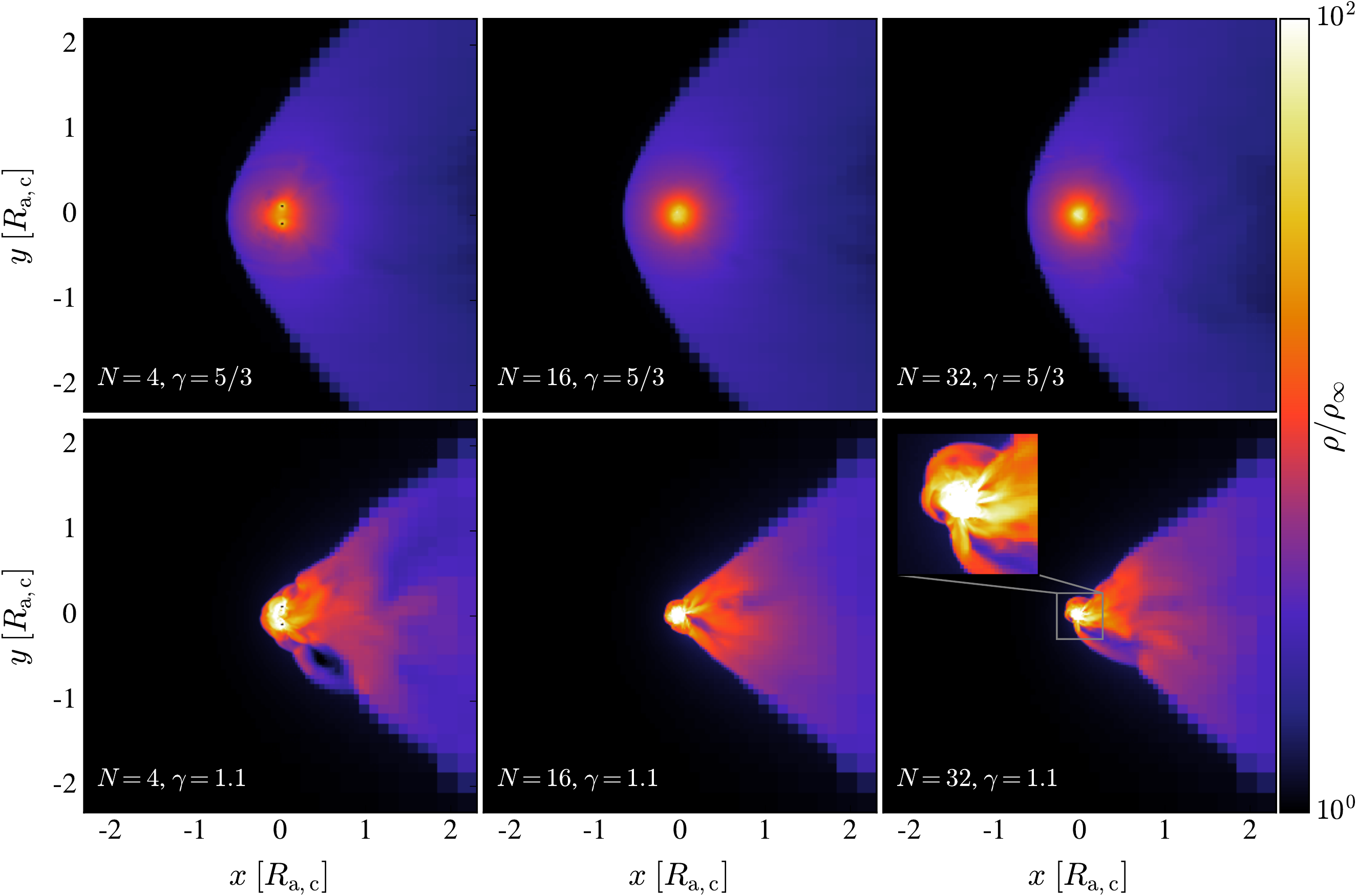}
\caption{The large-scale  structure of the flow is shown for $\gamma = 5/3$ and $\gamma = 1.1$ simulations with varying $N$. In all simulations $R_{\perp} = 0.5$ $R_{\rm a,{\star}}$. The density is depicted at time $32$ $R_{\rm a,c}/v_{\infty}$ in the $z=0$ plane. All adiabatic simulations exhibit a stable bow shock, as also shown in Figure~\ref{flow_structure_adiabatic}. All $\gamma = 1.1$ simulations, on the other hand, exhibit less stable flow. The ram pressure of the wind penetrates deep within the cluster's core and the higher compressibility of the flow leads to significantly larger density enhancements, consistent with studies of core potentials \citep{naiman_2009,naiman_2011}.\\}
\label{flow_structure_isothermal}
\end{figure*}

\subsection{The Role of Cooling on Accretion}
\label{subsec:results_eos}

Pressure support certainly plays a key role in BHL simulations \citep{2015ApJ...803...41M}. As a result of cooling, material can be compressed to varying degrees. In cases where the flow  can cool efficiently, the gas can be treated as nearly isothermal. The effects of radiative cooling are approximated here by having the gas evolve with an adiabatic constant $\gamma=1.1$, rather than with $\gamma = 5/3$ which models inefficient cooling. The effects of self-gravity of the gas are neglected. This is an adequate assumption for most clusters, for which the accumulated mass is less than the stellar mass. This assumption is certainly  not adequate for young star clusters in the process of formation \citep{2018arXiv181201615K}.

In simulations with a $\gamma = 1.1$ equation of state, the gas is much more compressible than in simulations with a $\gamma = 5/3$ equation of state. This has a dramatic effect on the morphology of the flow, as seen in Figure \ref{flow_structure_isothermal}, where we plot the density of the large-scale flow structure for both adiabatic and quasi-isothermal simulations  with $R_{\perp}/R_{\rm a,{\star}} = 0.5$.

\begin{figure}
\includegraphics[width=1.0\linewidth]{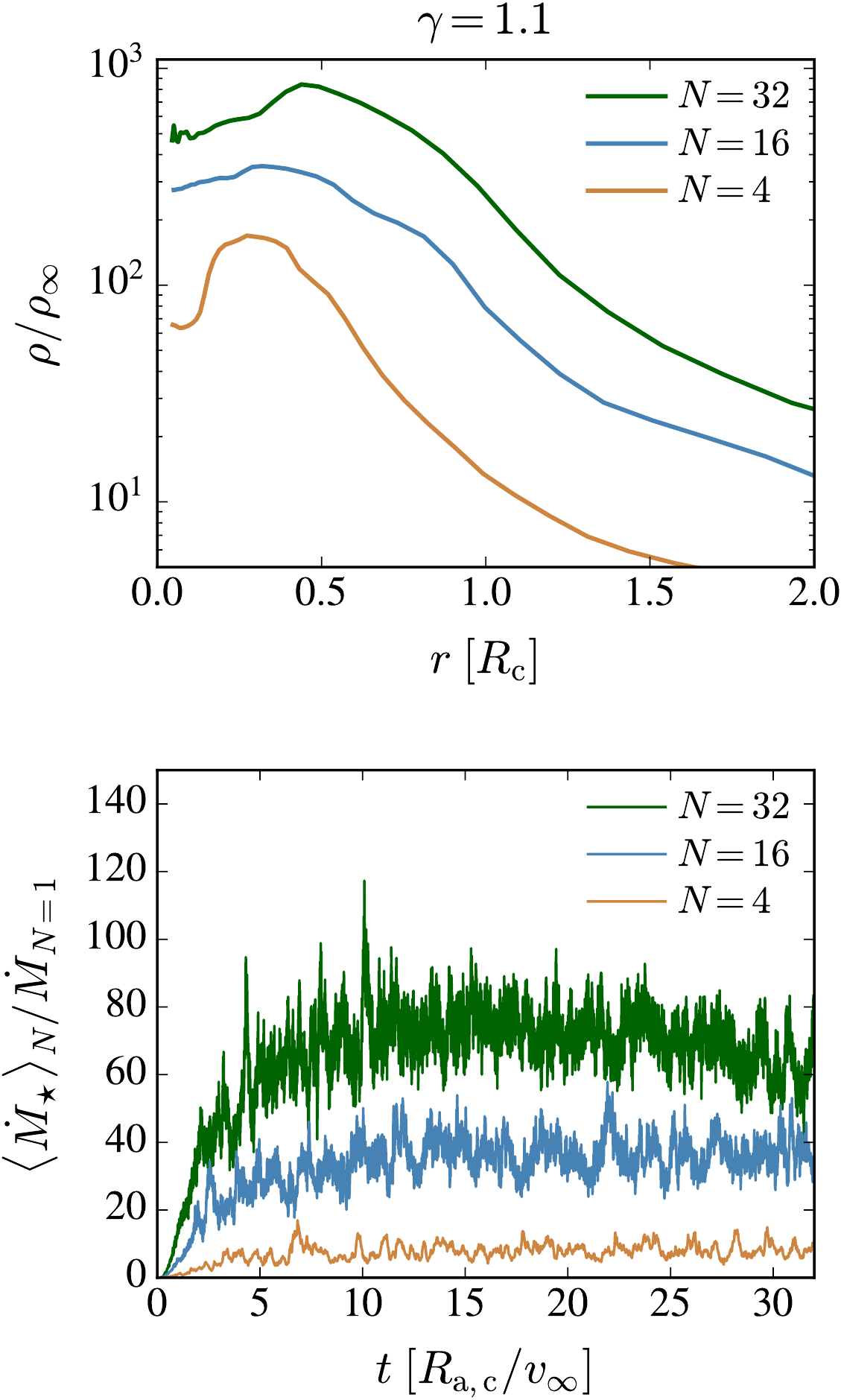}
\caption{Same as Figure~\ref{mdot_t_05rs_adiabatic} but for the $\gamma=1.1$ simulations. The number-averaged accretion rate is normalized  in the $\gamma=1.1$ case to $\dot{M}_{N=1} = 1.04$ $\dot{M}_{\rm HL,{\star}}$. As in the case of $\gamma=5/3$, the gas density and the individual accretion rates increase with $N$. The degree of variability of the accretion rate is  observed to increase with $N$, yet the characteristic temporal scale remains unchanged.}
\label{mdot_t_05rs_isothermal}
\end{figure}

In the adiabatic  simulations, corresponding to $\gamma=5/3$, a stable density enhancement is able to form, as  depicted  also in Figure~\ref{flow_structure_adiabatic} and  in agreement with studies of core potentials \citep{naiman_2011}.  In the $\gamma=1.1$ simulations, the enhanced compressibility of the flow allows for the effective buildup of mass at smaller scales and, as a result, the individual cluster members are able to accrete more efficiently. Adiabatic flows will instead rapidly heat the gas as it accumulates in the cluster potential, preventing the inflow of material to the core of the cluster at much lower threshold densities than in the $\gamma =1.1$ case.  The gas accumulated in a cluster with  $\gamma=5/3$ will be less centrally confined and less successful at insulating the accumulated gas from being advected by the incoming flow.  Although the high density region surrounding the cluster in the $\gamma = 5/3$ simulations is stable, it is by no means static.  This material is being constantly advected and replenished by material crossing the primary shock.  In the $\gamma = 1.1$ case, on the other hand, material penetrates deep into the potential of the individual accretors where it is captured before it can be advected downstream.

Our quasi-isothermal simulations also show the presence of instabilities within fast moving and highly compressible flow, similar to those seen in \citet{naiman_2011}. In highly compressible flows, modest pressure perturbations in  the shock region are able to forcefully produce sizable vorticity perturbations \citep{naiman_2011}. The differences in the flow structure are obvious when comparing the density maps shown in Figure~\ref{flow_structure_isothermal}, which show that the central density enhancements in adiabatic flows are
quasi-hydrostatic, while those for the $\gamma=1.1$ flows are not.

Figure~\ref{mdot_t_05rs_isothermal} compares the various simulated density profiles and individual accretion rates for the set of $\gamma=1.1$ simulations. These curves can be directly juxtaposed with those depicted  in Figure~\ref{mdot_t_05rs_adiabatic} for the  adiabatic cases.
Individual accretion rates are, as expected, much higher in the $\gamma = 1.1$ case. There is also a stronger dependence on $N$, with a corresponding increase in the degree of variability of $\nave\mdotstar$. Yet, the characteristic variability timescale remains similar across $N$ as it is primarily driven by $R_{\perp}/R_{\rm a,{\star}}$, which is here fixed to a value of $0.5$. 

\begin{figure}
\includegraphics[width=1.0\linewidth]{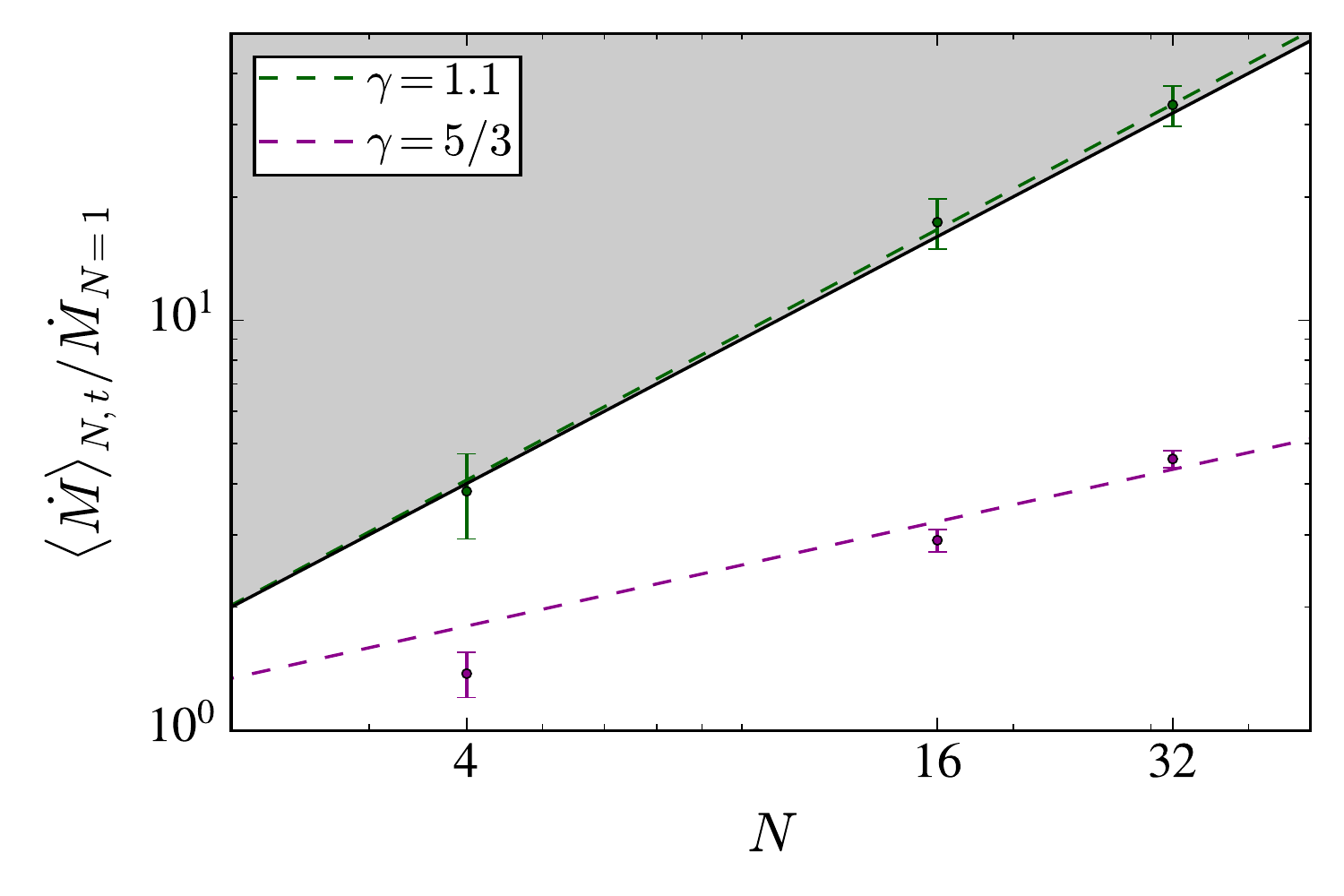}
\caption{The values of $\ntave\mdotstar$ in units of $\mdotone$ are plotted as a function of $N$ for $\gamma = 1.1$ and $5/3$. In all simulations $R_{\perp}/R_{\rm a,{\star}} = 0.5$. The simulation data for both equations of state are fitted to Equation~\ref{mdot_alpha} giving $\alpha = 1.01\pm0.03$ and $0.42\pm0.01$ for $\gamma = 1.1$ and $5/3$, respectively. In contrast to the adiabatic flow results presented in Figure~\ref{mdot_scale_ad}, the cluster as a whole can accrete at maximum efficiency when there is significant overlap between their accretion radii and cooling is efficient.}
\label{mdot_scale_it}
\end{figure}

In Figure \ref{mdot_scale_it}, we present the time- and number-averaged accretion rates as a function of $N$ for all $R_{\perp}/R_{\rm a,{\star}} = 0.5$ simulations, with the goal of  contrasting the adiabatic and quasi-isothermal results. Fitting our simulation data to Equation~\ref{eq:mdot_alpha_numerical}  we obtain  $\alpha = 1.01\pm 0.03$ for $\gamma=1.1$. For the quasi-isothermal equation of state, this indicates that the cluster as a whole can accrete at maximal efficiency. This is attributed to the heightened compressibility of the gas, which minimizes the advection of the material away from the individual accretors. 

\section{Discussion}
\label{sec:discussion}
We have shown that the structure of a supersonic flow around  discretized clusters can be characterized by a few essential parameters:  the individual accretion radius ($\rastar$), the mean separation between objects ($\rperp$), and the total number of objects ($N$). The relation between $\mdotstar$, $N$ and $\rperp/\rastar$ has been studied in previous sections by means of hydrodynamic simulations. The primary goals of this section are threefold;
\begin{itemize}
    \item To explore the validity of our results when considering variations in the accretion rate across cluster members,
    \item To present a summary of the salient lessons learned from our dimensionless calculations, and
    \item To discuss  how these lessons are applicable to relevant astrophysical settings.
\end{itemize}

\subsection{Dependence of $\mdotstar$ on Location}
So far, we have considered accretion rates averaged over all of the members in the cluster.  An important benefit of performing simulations of discretized clusters, though, is the positional information we have for accretion within the cluster. Each accretor is a probe of both the spatial distribution of the density within the cluster and the local flow properties (such as turbulent motions and local pressure gradients that may assist or inhibit accretion).  

\begin{figure}
\includegraphics[width=1.0\linewidth]{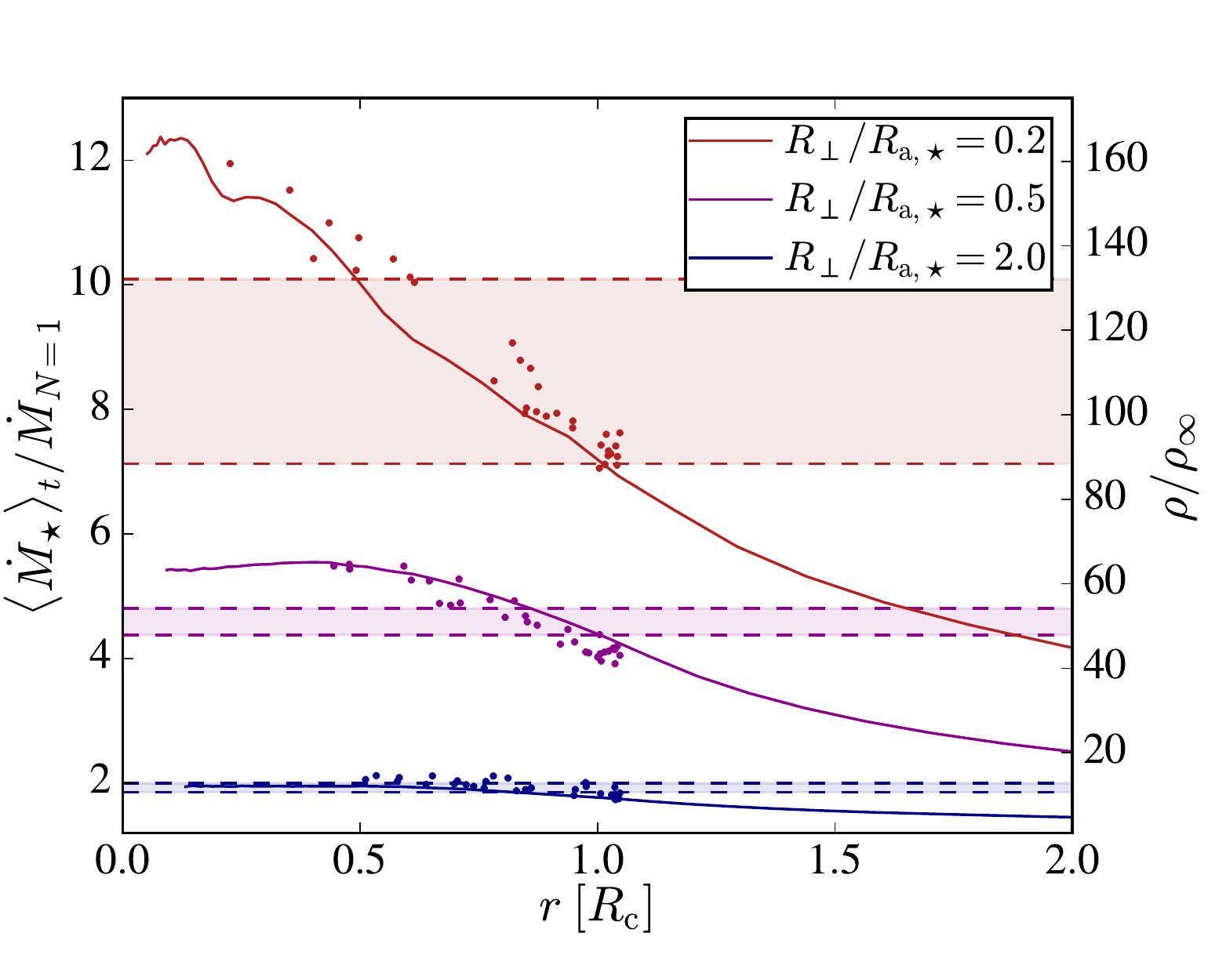}
\caption{The individual accretion rates of all  cluster members are plotted as a function of their radial distance from the cluster's center of mass. This is done  for the three simulations with $N = 32$ and $\gamma=5/3$. The individual points represent the time-averaged individual accretion rates of cluster members (left y-axis) and the solid lines represent the spatially-averaged density profiles (right y-axis) of each simulation. The shaded regions encompass the one sigma width of the time- and number-averaged accretion rate of each simulation (Figure \ref{mdot_scale_ad}). 
The scaling of the individual accretion rate with the corresponding density profile appears remarkably similar for different simulations, which in  turn depends sensitively on $\rperp/\rastar$.}
\label{structure_overplot}
\end{figure}

To study accretion as a function of position within a cluster, we plot in Figure~\ref{structure_overplot} the individual  mass accretion rates in three  $\gamma = 5/3$ simulations with $\rperp/\rastar = 0.2, 0.5,$ and 2.  For each simulation, the time-averaged accretion rate of all $N = 32$ accretors is plotted as well as their corresponding  radial distance (in units of $\rc$) from the  cluster's center of mass. The spatially-averaged radial density profiles are  over plotted  for comparison as  solid curves in Figure~\ref{structure_overplot}. Individual accretion rates are seen to systematically increase toward the center of the cluster and, as previously argued, roughly follow the corresponding density profile of the accumulated mass.

The correlation between $\ntave\mdotstar$  and $\rperp/\rastar$, reported in Figure~\ref{mdot_scale_ad}, is clearly preserved. That is,  $\tave\mdotstar$  is observed to systematically increase with decreasing  $\rperp/\rastar$ across the entire cluster. There is a good and bad side to this. On the negative side, it implies that one cannot be too specific about the exact value of $\alpha$, given that it depends on how the average is performed. On the positive side, it gives further credence to the validity  of Equation \ref{mdot_alpha} and the dependence of $\mdotstar$ on $\rperp/\rastar$. 

\subsection{Summary}
We have studied the conditions required for a  cluster of point masses to collectively accrete large amounts of gas from the external medium. We have shown that the ability of individual accretors to influence one another depends primarily on $\rperp / \rastar$ and $N$.  
When  $\rperp / \rastar \lesssim 1$, the accretion radii of the individual cluster members substantially overlap,  which permits them to accrete significantly more gas than they would in isolation. A key result of this study is the confirmation of the validity of the $\langle\dot{M_\star}\rangle = N^{\alpha}\dot{M}_{\rm HL,\star}$ relation and the exploration of the $\alpha$ parameter dependence with $\rperp/\rastar$ and the equation of state of the gas. The specific salient  findings  may be summarized as follows:

\begin{itemize}
    \item When the separation between  accretors  is  much  larger  than  their  accretion  radii, each accretor can be considered to accrete independently. That is, $\alpha \approx 0$ when $\rperp \gg \rastar$.
    \item When the accretion radii of the cluster members  substantially  overlap, each of them can accrete significantly more than they would in isolation. As a result, $\alpha$ steepens as $\rperp/\rastar$ decreases although  it remains sub-linear ($\alpha\lesssim 0.6$).  In the adiabatic simulations, we find $\alpha$  to be $0.18\pm0.01$, $0.42\pm0.01$ and $0.62\pm0.04$ for  $R_{\perp}/R_{\rm a,{\star}} = 2$, $0.5$ and $0.2$  respectively (Figure \ref{mdot_scale_ad}). 
    \item When  $\gamma = 1.1$,  the  enhanced  compressibility  of  the  flow  allows  for  an effective buildup of gas in the cluster's core, which permits the embedded objects to accrete at maximum efficiency provided that $\rastar/\rperp \lesssim 1$. In simulations with $\gamma = 1.1$ we find $\alpha = 1.01\pm0.03$ for $\rperp / \rastar = 0.5$.  In contrast, for the adiabatic case we find $0.42\pm0.01$ for $\rperp / \rastar  = 0.5$ (Figure \ref{mdot_scale_it}). 
\end{itemize}

\subsection{Astrophysical Relevance}\label{subsec:astro}
To circumvent computational resolution constraints, the dimensionless calculations presented in this paper have been constructed under the assumptions that the cluster members are distributed uniformly and that $\rac$ remains unchanged.   These conditions imply, as argued in Section~\ref{subsec:sp}, that the velocity dispersion of our simulated clusters scales as $\sigma_{\rm c}\propto N^{1/3}$. This sets the scaling of the characteristic velocity with mass for our simulated clusters. 

The classical BHL treatment in clusters fails to provide an accurate description  when $\vinf /\sigma_{\rm c} \lesssim 1$. This is equivalent to the critical  condition $\rc/\rac \lesssim 1$, which we have reviewed in  Section~\ref{sec:background} and was originally outlined in \citet{naiman_2011}.
In such cases,  the collective potential is able to alter the properties of the ambient gas, allowing the system to accumulate subsonic material within the cluster and enabling the cluster members to accrete at enhanced rates. 

The conditions and associated scaling relations derived in this paper  are largely applicable to systems that collectively accrete ($\vinf/ \sigma_{\rm c} \lesssim 1$) and are well described by a core gravitational potential with $\sigma_{\rm c}\propto N^{1/3}$.  The evolution of $\rc$ for a stellar system with a realistic mass function in which the stellar members evolve \citep{gurkan_2004,2010MNRAS.408L..16G} and are also influenced by the tidal field \citep{lee_goodman_1995} of the host galaxy is rather complex. Yet, broadly speaking, globular clusters can be relatively well described by potentials with $\sigma_{\rm c}\propto N^{1/2}$ \citep{2005ApJ...627..203H}, while  young stellar clusters roughly  follow  $\sigma_{\rm c}\propto N^{1/4}$ \citep{adams_2006,pfalzner_2011,2016A&A...586A..68P}.  Because of this, the scaling laws derived  from our dimensionless calculations  are broadly applicable to these systems. It is important to note that our simulations assume, for simplicity, that the individual cluster members are static. This assumption neglects the heating rates caused by the motion of the stars, which can induce shocks when the cluster velocity dispersion is larger than $\vinf$. For the cluster velocity dispersions discussed here, this heating rate might be sizable and thus should be included in future work.

The consequences of enhanced mass accumulation in clusters fall into two broad categories. First, if the gas cools efficiently, then  star formation might be triggered. This could give rise to new stellar members in the cluster.  Second, if the gas accretes efficiently onto the cluster, the state of the embedded stellar systems can be  altered.

Effective mass accumulation in star clusters can only be sustained if
they move with a relatively slow ($\mach\lesssim 2$) velocity  with respect to the ambient gas \citep{naiman_2011}.  In the Milky Way, globular clusters are seen to move at highly supersonic ($\mach\approx 5-10$) velocities with respect to the ambient gas \citep{dauphole_1996,1997AJ....114.1014D}. What is more, the gas within the galactic halo is sufficiently hot such that $\rac \ll \rc$. Effective accretion might thus only occur in these older stellar systems if they have low-inclinations and low-eccentricities \citep{lin_2007}. 

The environments surrounding young clusters are by contrast  better suited  for  large mass accumulation. This is the case for young clusters embedded in galactic disks \citep{2018arXiv181201615K} or those found in merging galaxies \citep{2007ApJ...668..168G}. For example, observations of several clusters in the Antennae show cluster velocities relative to the gas that are comparable to the  velocity dispersions of the embedded clusters \citep{2005AJ....130.2104W}, implying  that $\rac \gtrsim \rc$. Observations at infrared wavelengths also show that large amounts of dust are present in these systems \citep{2005ApJ...635..280B}, which justifies the quasi-isothermal assumption.

The gas accumulated in these clusters might be able to cool efficiently, and could  fragment and form stars provided that the density is high enough. More detailed simulations that take into account realistic cooling and self-gravity of the gas are, however, needed before inferences  can be made about the relevance of our calculations to clusters showing subsequent episodes of star formation \citep[e.g.,][]{pflamm_kroupa_2009,2011ApJ...726...36C}. 

Even if the density is not above the star formation threshold, the resulting accretion rates in these systems (which are drastically enhanced)  could change the state of the cluster members. For example, the accretion of gas in metal poor clusters could potentially alter the  birth metallicity  of the stars \citep{talbot_newman_1977, yoshii_1981, 2017MNRAS.469.4012S} and, more generally,  lead to abundance variations \citep{2004ARA&A..42..385G,dercole_2010}. Moderately high accretion rates in the white dwarf cluster population could, for instance,  result in enhanced novae rates \citep{2007ApJ...663.1269N, lin_2007,naiman_2011} and could unexpectedly alter the cooling sequence \citep{1952MNRAS.112..583M}. 
Large density accumulation of gas could also be accompanied by enhanced mass supply to the central populations of LIGO black holes expected to reside in these systems \citep{2015PhRvL.115e1101R}. This could result  in a sizable black hole accretion luminosity, which could produce an ultraluminous, compact X-ray source, similar to those that  have been preferentially found in association with young clusters \citep{2001ApJ...554.1035F,2007ApJ...661..135Z,2009A&A...501..445C}. The detection or non-detection  of such ultraluminous X-ray sources could offer strong constraints on the population of LIGO black holes  and the nature of the stellar clusters that host them.

\begin{acknowledgements}
The notions expressed in this work have grown out of several exchanges with D. Lee, D.~Lin, M.~MacLeod and J.~Naiman. We are indebted to them for guidance and encouragement. We acknowledge support from the DNRF (Niels Bohr Professor), NASA and the Packard Foundation. The authors thank the Niels Bohr Institute for its hospitality while part of this work was completed, and the Kavli Foundation and the DNRF for supporting the 2017 Kavli Summer Program. The simulations presented in this work were performed using software produced by the Flash Center for Computational Science at the University of Chicago, which was visualized using code managed by the yt Project. 
\end{acknowledgements}

\bibliographystyle{aasjournal}
\bibliography{clusterBHL_references}

\end{document}